# Strain Relaxation via Phase Transformation in SrSnO$_3$

Tristan K Truttmann[1,a], Fengdeng Liu[1], Javier Garcia Barriocanal[2], Richard D. James[3], and Bharat Jalan[1,a]

[1]Department of Chemical Engineering and Materials Science, University of Minnesota – Twin Cities, Minneapolis, MN 55455, USA
[2]Characterization Facility, University of Minnesota, Minneapolis, Minnesota – Twin Cities, 55455, USA
[3]Department of Aerospace Engineering and Mechanics, University of Minnesota – Twin Cities, Minneapolis, MN 55455, USA.

a) trutt009@umn.edu, bjalan@umn.edu

**Abstract**

SrSnO$_3$ (SSO) is an emerging ultra-wide bandgap (UWBG) semiconductor with potential for high-power applications. In-plane compressive strain was recently shown to stabilize the high-temperature tetragonal phase of SSO at room temperature (RT) which exists at $T \geq 1062$ K in bulk. Here, we report on the study of strain relaxation in epitaxial, tetragonal phase of Nd-doped SSO films grown on GdScO$_3$ (110) (GSO) substrates using radical-based hybrid molecular beam epitaxy. The thinnest SSO film (thickness, $t = 12$ nm) yielded a fully coherent tetragonal phase at RT. At 12 nm $< t <$ 110 nm, the tetragonal phase first transformed into orthorhombic phase and then at $t \geq 110$ nm, the orthorhombic phase began to relax by forming misfit dislocations. Remarkably, the tetragonal phase remained fully coherent until it completely transformed into the orthorhombic phase. Using thickness- and temperature-dependent electronic transport measurements, we discuss the important roles of the surface, phase coexistence, and misfit dislocations on carrier density and mobility in Nd-doped SSO. This study provides unprecedented insights into the strain relaxation behavior and its consequences for electronic transport in doped SSO with implications in the development of high-power electronic devices.

The family of perovskite alkaline-earth stannates has demonstrated room-temperature (RT) electron mobilities as high as 320 cm$^2$V$^{-1}$s$^{-1}$ in doped BaSnO$_3$ (BSO) bulk crystals[1], a tunable bandgap (indirect) from ~ 3 eV in BSO to ~ 4.1 eV in SrSnO$_3$ (SSO)[2-4], and even *p*-type doping.[5,6] In contrast to bulk single crystals, thin films of BSO have only achieved electron mobilities of 182 cm$^2$V$^{-1}$s$^{-1}$ at a carrier density of 1.2 × 10$^{20}$ cm$^{-3}$.[7] This discrepancy in electron mobility between BSO films and bulk crystals is attributed largely to the presence of threading dislocations in thin films owing to the lack of commercially available lattice-matched substrates.[7-9] There are ongoing efforts toward the development of homoepitaxy[10], buffer layers[8,11] and lattice-matched substrates[6,12,13] to reduce the density of threading dislocations in BSO, but none of these efforts have yet resulted in electron mobilities that rival those in bulk single-crystals. SSO is a non-cubic member of the stannate family. Similar to BSO, SSO's conduction band is derived from Sn-5*s* orbitals and consequentially has a low electron effective mass.[14-16] SSO can also be doped *n*-type with rare-earth elements and has demonstrated room-temperature electron mobilities as high as 70 cm$^2$V$^{-1}$s$^{-1}$ in thin films.[14, 17-20] Unlike BSO, however, SSO has a wider band gap (~ 4.1 eV)[2-4] which makes it more attractive as an ultra-wide bandgap (UWBG) semiconductor for power device applications.[21] Importantly, SSO can also be grown coherently on commercially available substrates due to its smaller lattice parameter which not only offers a path to avoid the threading dislocations which have plagued BSO films, but also opens the door to strain engineering.[14, 17] Earlier work from our group has shown that compressive epitaxial strain can stabilize the high-temperature tetragonal polymorph (T) of bulk SSO to below RT, more than 700 K below its stability range in bulk.[14] Films grown under no strain or tensile strain adopt the RT orthorhombic polymorph (O) as illustrated in Fig. 1. However, it is yet to be understood how these phases evolve with increasing film thickness. For instance, will strain-stabilized tetragonal SSO on GSO (110) undergo a phase transition or will it form misfit dislocations if the film thickness is increased? It is also conceivable that the elastic strain energy can be accommodated through oxygen octahedral rotation/tilt[22], polarization[23], ferroelastic domain formation[24], and/or structural phase transition.[25, 26]

In this paper, we investigate strain relaxation in the strain-stabilized tetragonal phase of Nd-doped SSO films grown on GSO (110) substrates. We find that phase transformation ($T \rightarrow O$) *precedes* strain relaxation via the formation of misfit dislocations with increasing film thickness. A significant increase in carrier density from $1.3 \times 10^{18}$ to $1.3 \times 10^{20}$ cm$^{-3}$ was discovered accompanied by an increase in mobility from 14 to 73 cm$^2$V$^{-1}$s$^{-1}$ between 12 nm and 330 nm. Both electron density and mobility remained unchanged during strain relaxation via misfit dislocations at $t \geq 110$ nm suggesting dislocations in SSO films *may* not be charged unlike in BSO films. Surprisingly, we found a conduction dead layer of 12 nm, below which films remain insulating despite high doping concentration, $\sim 1.3 \times 10^{20}$ cm$^{-3}$. We discuss possible mechanisms for the existence of the dead layer.

Samples were grown using a radical-based hybrid MBE approach.[14, 27, 28] A brief description of this approach is included here. All films were grown in an EVO 50 MBE System (Omicron, Germany). Substrates were heated to 950 °C (thermocouple temperature). Before growth, the substrates were cleaned for 20 minutes using radio-frequency (RF) oxygen plasma (Mantis, UK) operating at 250 W and at oxygen pressure $5 \times 10^{-6}$ Torr. Strontium (Sr) was supplied from a thermal effusion cell with a beam-equivalent pressure (BEP) of $2.4 \times 10^{-8}$ Torr. Tin was supplied via a gas injector (E-Science inc.) using a radical-forming chemical precursor hexamethylditin (HMDT) at a BEP of $2 \times 10^{-6}$ Torr. All films were grown in the presence of oxygen plasma operating at 250 W and at oxygen pressure of $5 \times 10^{-6}$ Torr. These growth conditions yielded a growth rate of 55 nm/hour. For electrical measurements, films were doped *n*-type using neodymium (Nd). Nd was supplied from a thermal effusion cell operating at a fixed temperature of 940 °C.

Films were characterized using atomic force microscopy (AFM), high-resolution X-ray diffraction (HRXRD), and Van der Pauw (VdP) Hall measurements. AFMs were collected in contact mode using a Nanoscope V Multimode 8 (Bruker, Germany). HRXRD coupled scans and rocking curves were collected with an X'Pert Pro thin film diffractometer (PANalytical, Netherlands) equipped with a Cu parabolic mirror and Ge 4-bounce monochromator. Lattice parameters were extracted from reciprocal space maps (RSMs), which were collected with a SmartLab XE thin film diffractometer (Rigaku, Japan) collected with parallel-beam optics, a Ge 2-bounce monochromator, and HyPix-3000 2D Detector. Electrical measurements were performed in a physical property measurement system (DynaCool, Quantum Design, USA) using a VdP configuration. Indium was used to make ohmic metal contacts. Magnetic field and temperature were varied between ± 9 T and 1.8 K - 300 K respectively.

Fig. 2a shows on-axis high-resolution 2θ-ω coupled scans of $t$ nm Nd-doped SSO/10 nm SSO/GSO (110) where $t$ was varied between 12 and 330 nm. The film at $t$ = 12 nm revealed an expanded out-of-plane lattice parameter of 4.117 ± 0.002 Å in agreement with the strain-stabilized tetragonal SSO polymorph on GSO (110).[14] This film also showed finite-size thickness fringes consistent with high structural quality and smooth surface morphology. With increasing $t$, films ($t$ = 26 nm and $t$ = 55 nm) showed an additional XRD peak corresponding to an orthorhombic phase.[14] It is noted that although the two peaks for $t$ = 26 nm are not apparent in the $(002)_{pc}$ region of Fig. 2a, the distinction is clearly visible in the $(103)_{pc}$ RSM of Fig. 3 (discussed below). With further increasing $t$, the tetragonal phase vanishes whereas the peak corresponding to the orthorhombic phase shifts towards higher 2θ values. Fig. 2b shows the corresponding rocking curves as a function of $t$. These rocking curves can be described as a linear combination of two Gaussians (a narrow and a broad component).[17] This behavior is commonly seen in thin films where the narrow component reflects the low degree of structural disorder in coherent films, and the broad component reflects the disorder generated during strain relaxation.[29] We therefore plot $\frac{I_{broad}}{I_{broad}+I_{narrow}}$ as a function of $t$ in Fig. 2c to investigate the evolution of structural disorder with film thickness. Here, $I_{broad}$ and $I_{narrow}$ are the peak intensity of the broad and narrow Gaussian components, respectively. The value of $\frac{I_{broad}}{I_{broad}+I_{narrow}}$ was found to increase with $t$ reaching a constant value of 1 for $t \geq 110$ nm suggesting higher structural disorder for thicker films. To this end, it is apparent that strain relaxation is initiated via phase transition and that it is accompanied by increasing structural disorder. It remains unclear, however, whether or not these phases relax via forming misfit dislocations at substrate/film interface.

In order to investigate this question, we measured both in-plane and out-of-plane lattice parameters as a function of $t$. Fig. 3 shows asymmetric RSMs around the $(103)_{pc}$ reflection (top panels) and symmetric RSMs around the $(002)_{pc}$ reflection (bottom panels) for $t$ nm Nd-doped SSO/10 nm SSO/GSO (110) as a function of $t$. Consistent with the coupled scan, RSMs again confirmed T → O transformation followed by a change in the lattice parameters. To illustrate this point, we show in Fig. 4a measured in-plane ($a_{ip}$) and out-of-plane ($a_{op}$) lattice parameters of both phases determined from the analysis of the RSMs as a function of $t$. The tetragonal phase is represented using diamond symbols whereas square symbols denote the orthorhombic phase. Fig. 4a shows the emergence of the orthorhombic phase for $t \geq 26$ nm, and it reveals that both orthorhombic and tetragonal phases remain fully coherent to the GSO substrate up to $t = 55$ nm. Despite no change in $a_{ip}$, $a_{op}$ of the orthorhombic phase was found to decrease whereas $a_{op}$ of the tetragonal phase remained unchanged. At $t = 110$ nm, a partially relaxed orthorhombic phase was observed with no measurable tetragonal phase suggesting strain relaxation via the $T \rightarrow O$ transformation completes between 55 nm and 110 nm. The film then begins to relax by forming misfit dislocation accompanied by a change in lattice parameters reaching a bulk value ~ 4.035 Å at $t = 330$ nm.[14, 30]

We now discuss these lattice parameters in detail. At $t \geq 110$ nm, the orthorhombic phase undergoes a decrease in $a_{op}$ accompanied by an increase in $a_{in}$. This is an expected behavior from strain (compressive) relaxation owing to the formation of misfit dislocations. However, it may not be self-evident why the $T \rightarrow O$ transformation occurs when $a_{in}$ remains unchanged. This is contrary to expectations, as the constraint on $a_{ip}$ via coherent strain is understood to be the cause of the strain-stabilized tetragonal phase. But here, it turns out that strain stabilization perishes while $a_{ip}$ is still at its constrained value. Additionally, it remained puzzling why the $T \rightarrow O$ transformation (for 26 nm $\leq t \leq$ 55 nm) is accompanied by a decrease in $a_{op}$ only for the orthorhombic phase whereas $a_{in}$ for both phases remains unchanged. To provide a qualitative explanation for this observation, we refer to Fig. 1 and remind the reader that an epilayer undergoing symmetry ($T \rightarrow O$) relaxation is characterized by a coexistence of these two phases separated by coherent phase boundaries.[31] When the epilayer is predominantly tetragonal with only small volume fractions of orthorhombic phase (such as $t = 26$ nm as shown in Fig. 3c), all orthorhombic phase is directly adjacent to its tetragonal counterpart.[31] Therefore, $a_{op}$ of the orthorhombic fraction is pinned to that of tetragonal phase via a *laterally directed coherent strain*. At larger thicknesses when less tetragonal phase is present ($t = 55$ nm as shown in Fig. 3d), the orthorhombic phase is separated from the tetragonal regions by additional orthorhombic phase. In this case, laterally directed coherent strain relaxes, resulting in a decrease in $a_{op}$ for the orthorhombic phase. But why is the orthorhombic lattice parameter influenced by the tetragonal phase, and not vice-versa? We explain this again by referring to Fig. 1, where we see that $a_{op}$ for the tetragonal phase is fully defined by the length of Sn-O bonds which are colinear and oriented in the out-of-plane direction. For the orthorhombic phase, however, $a_{op}$ is a function of both Sn-O bond distance *and* the octahedral rotation angles. Therefore, the orthorhombic phase can easily accommodate larger $a_{op}$ by simply decreasing the angles of the $a^+$ or $b^-$ rotations. Future investigations using transmission electron microscopy should be directed to investigate the laterally directed coherent strain relaxation process.

Finally, we discuss the effect of strain relaxation on electronic transport. Figures 4b and 4c show room temperature electron density ($n_{300K}$) and mobility ($\mu_{300K}$) as a function of *t*. Both $n_{300K}$ and $\mu_{300K}$ were found to increase from $1.3 \times 10^{18}$ to $1.3 \times 10^{20}$ cm$^{-3}$ and from 14 to 67 cm$^2$V$^{-1}$s$^{-1}$, respectively, between 12 nm and 110 nm and then become approximately independent of film thicknesses. Coincidently, this is the same thickness range in which strain relaxation was found to occur via the $T \rightarrow O$ transformation raising the question: what are the effects of phase fraction and phase boundary on electron density and mobility? Additionally, no measurable change in $n_{300K}$ and $\mu_{300K}$ was observed for films with $t \geq 110$ nm questioning the nature of dislocations in SSO films. To put this into context, threading dislocations in doped BSO are charged and have been shown to compensate carrier density while lowering the mobility. In contrast, strain relaxation via the formation of dislocations (for $t \geq 110$ nm) did not yield any observable change in carrier density and mobility, suggesting dislocations in SSO may not be charged unlike in BSO. Additional work is needed to determine the composition of dislocation cores in doped SSO films and to establish the relationship between local composition and electronic structure. Future TEM work will be directed to investigate dislocations in doped SSO films. However, to investigate the increase in carrier density and mobility between 12 nm and 110 nm, we show in Fig. S1 the $n_{2D}$ as a function of $t$. In an uncompensated semiconductor, $n_{2D}$ vs. $t$ should follow a straight line passing through the origin where the slope of the line yields $n_{3D}$. As expected, our experimental data showed a linear behavior but with a finite $x$-intercept as $n_{2D} \rightarrow 0$. The slope yielded a 3D density of $1.4 \times 10^{20}$ cm$^{-3}$ consistent with the expected donor density based on the doping calibration. A finite $x$-intercept, however, suggests a conduction dead layer thickness of ~ 11.8 nm i.e. the film thickness over which electrons are fully compensated and do not contribute to conduction. A cause of the conduction dead layer in semiconductors can be the depletion effect from band bending due to charges at the surface. However, the 11.8 nm dead layer in this study would be extraordinarily large given our high donor density of $N_d = 1.4 \times 10^{20}$ cm$^{-3}$. In fact, assuming a relative permittivity value for SSO of $\varepsilon_r = 15$, this would correspond to a built-in surface potential of 12 V, which is inconceivable considering this would invert the valence band maximum well above the Fermi level at the surface. Therefore, the conduction dead layer in SSO cannot be explained by surface depletion alone. Rather, we hypothesize that it is due to the strong localization effect owing to surface disorder in thin films. To this end, we performed temperature-dependent resistivity and Hall measurements of films as a function of $t$ as shown in Figs 4(d-f) revealing degenerate semiconductor metallic behavior for films $t \geq 26$ nm between 1.8 K and 400 K. However, the thinnest film, $t = 12$ nm yielded insulating behavior indicating the important role of strong localization in thinner films where $t$ only slightly exceeds the conduction dead layer thickness. It is expected that an undoped capping layer may help mitigate this problem. Future study will be directed to investigate this effect.

In summary, we have performed a systematic study of strain relaxation in the strain-stabilized tetragonal phase of Nd-doped SSO films. It was found that strain relaxation occurs via $T \rightarrow O$ phase transformation followed by relaxation via misfit dislocations. We also reveal various subtleties of the strain relaxation process including how the volume fraction of the two phases affect $a_{\text{op}}$ of the orthorhombic phase via laterally directed coherent strain. Finally, we show that Nd-doped SSO has a conduction dead layer which cannot be explained by band bending alone, and we propose surface disorder-driven localization as a potential explanation for the dead layer. This study provides an important step forward utilizing SSO as a new UWBG perovskite semiconductor for the development of high-power electronic devices.

**Acknowledgements**

This work was supported by the Air Force Office of Scientific Research (AFOSR) through Grant No. FA9550-19-1-0245 and through NSF DMR-1741801. The work also benefitted from the Norwegian Centennial Chair Program (NOCC), and a Vannevar Bush Faculty Fellowship. Parts of this work were carried out at the Minnesota Nano Center and Characterization Facility, University of Minnesota, which receives partial support from NSF through the MRSEC program under Award Number DMR-2011401.

**Figures (Color Online):**

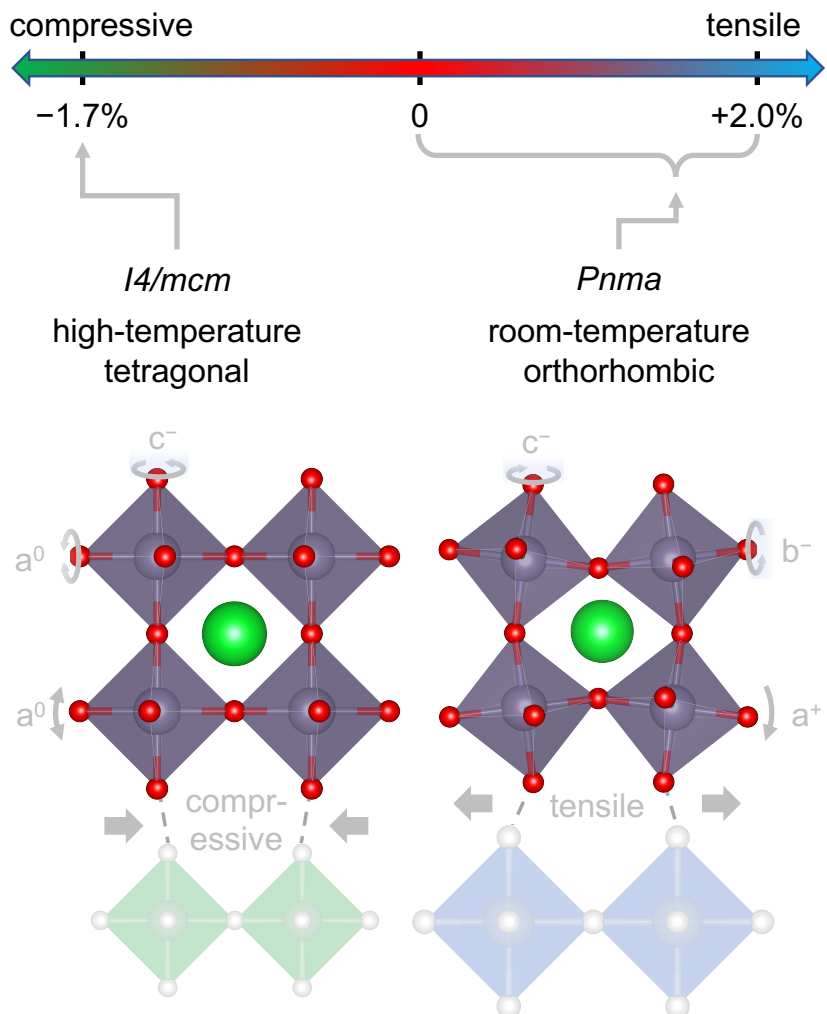

**Fig. 1 Strain stabilization of SSO.** Schematic illustrating strain stabilization of SSO under compressive and tensile strain. Compressive strain stabilizes the tetragonal phase (*I4/mcm* with $a^0a^0c^+$ tilt pattern) whereas no strain and tensile strain tends to stabilize a room-temperature orthorhombic phase (*Pnma* with $a^+b^-c^-$ tilt pattern).

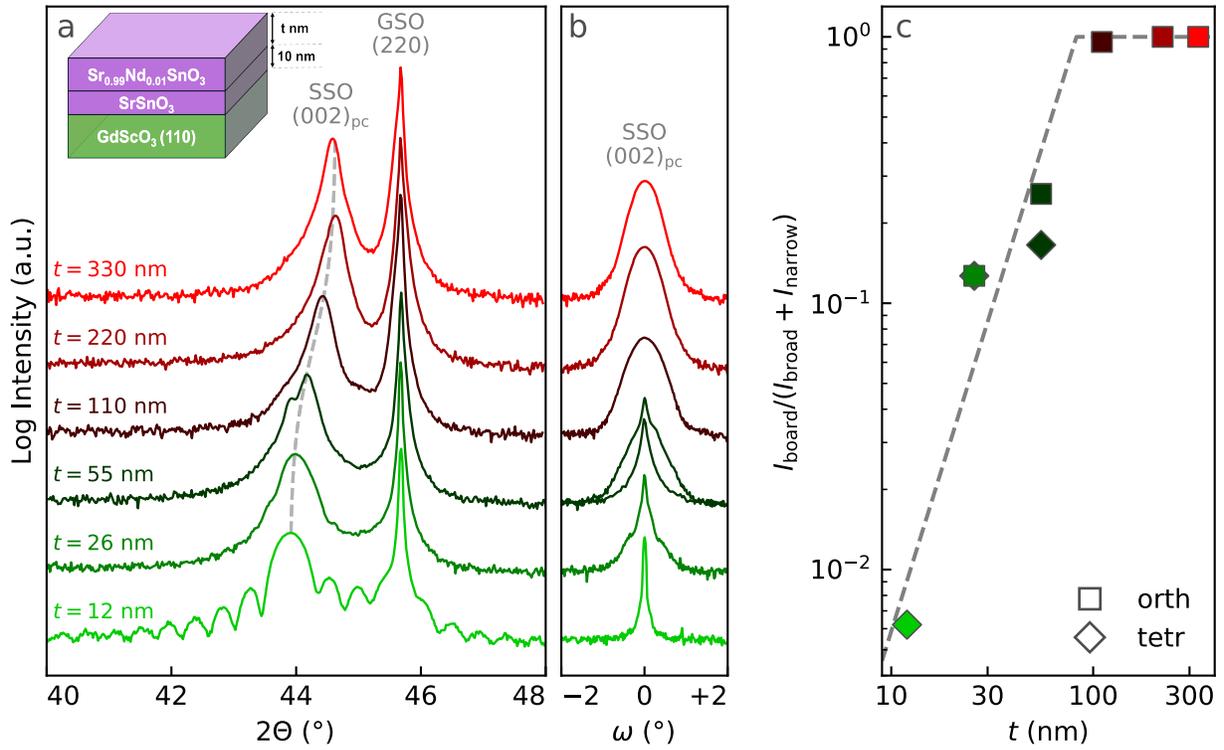

**Fig. 2 High-resolution X-ray diffraction.** Coupled scans (**a**), and rocking curves (**b**) of $t$ nm Nd-doped SSO/10 nm SSO/GSO. **c** $\frac{I_{broad}}{I_{broad}+I_{narrow}}$ as a function of $t$, where $I_{broad}$ and $I_{narrow}$ are peak height of the broad and narrow Gaussian components, respectively.

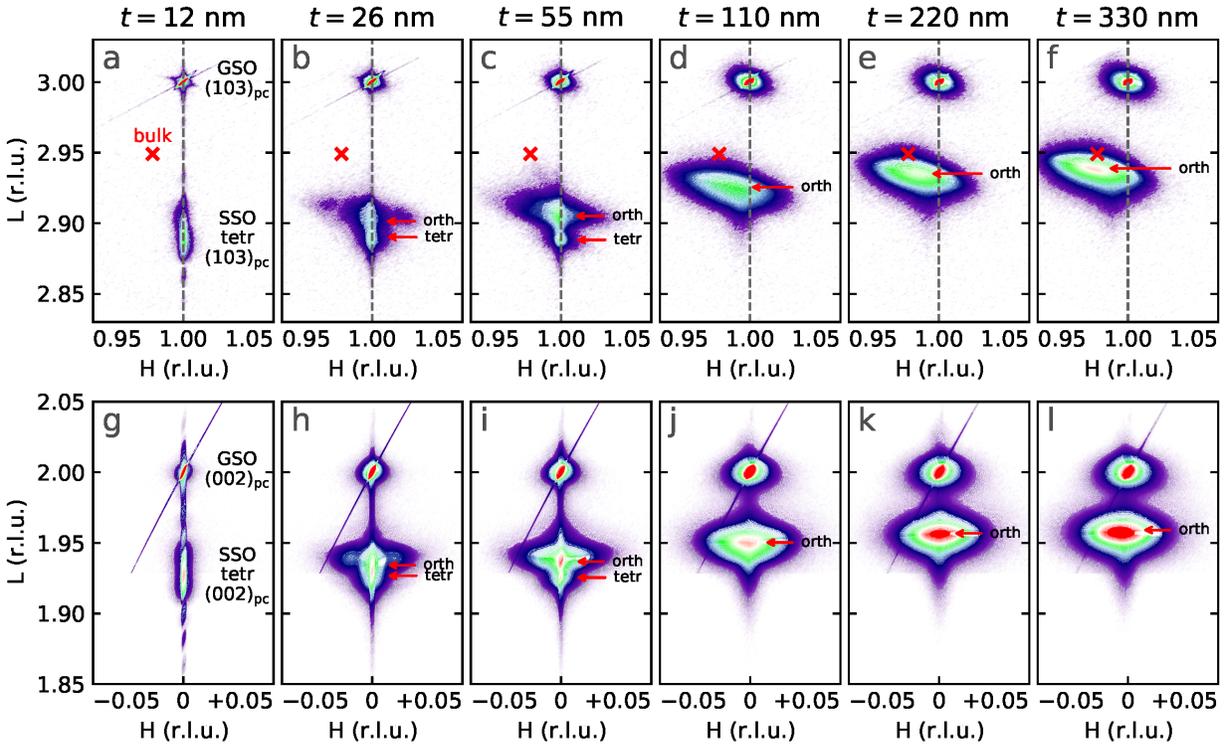

**Fig. 3 Thickness-Dependent Reciprocal Space Maps. a-f** Asymmetric RSMs around the $(103)_{pc}$ reflection, and (**g-l**) symmetric RSMs around the $(002)_{pc}$ reflection for *t* nm Nd-doped SSO/10 nm SSO/GSO (110) as a function of *t*. Red arrow indicates Bragg peaks corresponding to two phases present in the film. The cross symbol indicates position in reciprocal space corresponding to the fully-relaxed lattice parameter of SSO.

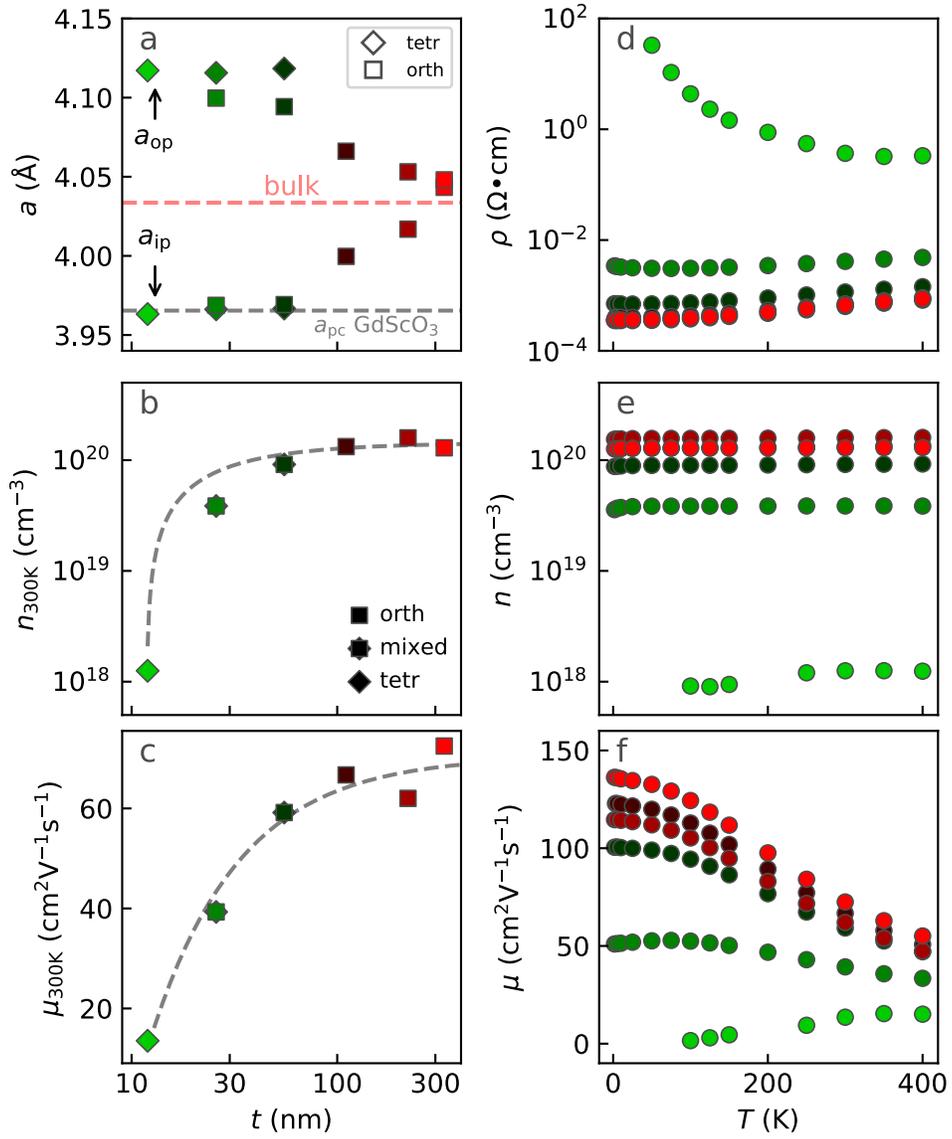

**Fig. 4 Lattice Parameters and Electronic Transport.** (**a**) In-plane ($a_{ip}$) and out of plane lattice parameter ($a_{op}$) of $t$ nm Nd-doped SSO/10 nm SSO/GSO (110) as a function of $t$. (**b-c**) Room-temperature 3D electron density $n_{300K}$, and mobility $\mu_{300K}$ versus $t$. (**d-f**) resistivity ($\rho$), $n_{3D}$ and $\mu$ of these films respectively as a function of temperature.

# Strain Relaxation via Phase Transformation in SrSnO$_3$

Tristan K Truttmann[1,a], Fengdeng Liu[1], Javier Garcia Barriocanal[2], Richards D. James[3], and Bharat Jalan[1,a]

[1]Department of Chemical Engineering and Materials Science, University of Minnesota – Twin Cities, Minneapolis, MN 55455, USA
[2]Characterization Facility, University of Minnesota, Minneapolis, Minnesota – Twin Cities, 55455, USA
[3]Department of Aerospace Engineering and Mechanics, University of Minnesota – Twin Cities, Minneapolis, MN 55455, USA.

a) trutt009@umn.edu, bjalan@umn.edu

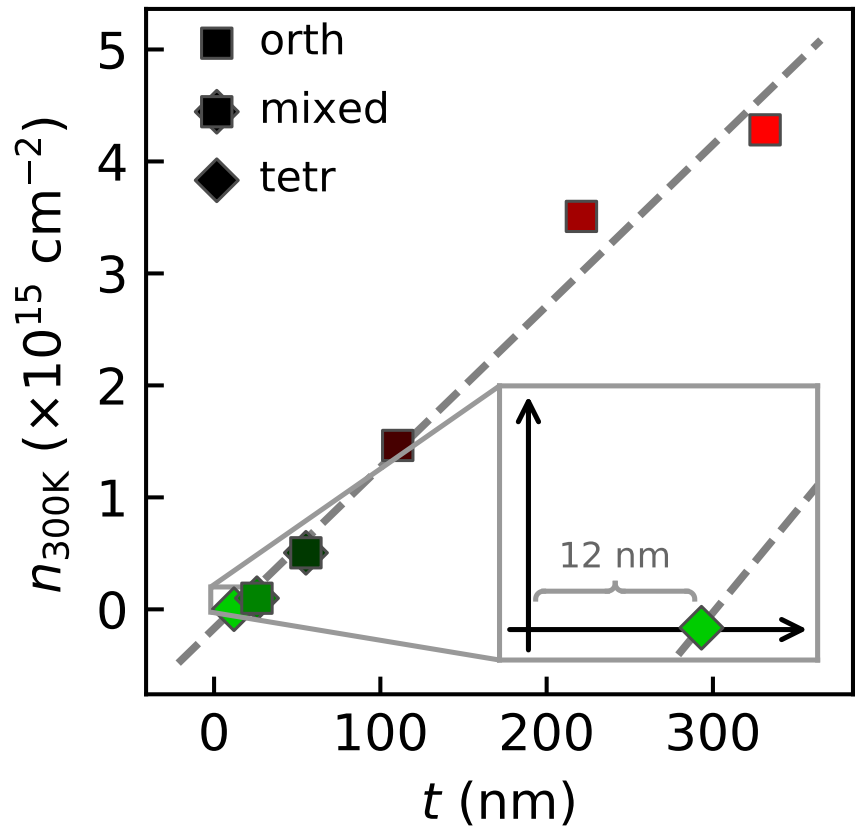

**Figure 1:** Sheet carrier density as a function of film thickness, $t$. Dashed line shows a linear fit to the data. Inset shows a zoom-in figure revealing $x$-intercept of ~ 11.6 nm as $n_{300K} \rightarrow 0$